\documentstyle[aps,twocolumn,psfig]{revtex}
\tightenlines
\begin{document}
\draft
\preprint{DUKE-TH-96-126}
\title{SOFT ELECTROMAGNETIC RADIATIONS FROM RELATIVISTIC HEAVY ION
COLLISIONS}
\author{Dipali Pal, Pradip Kumar Roy, Sourav Sarkar, Dinesh Kumar
Srivastava}
\address{Variable Energy Cyclotron Centre, 1/AF Bidhan Nagar, Calcutta
700 064}
\author{Bikash Sinha}
\address{Variable Energy Cyclotron Centre, 1/AF Bidhan Nagar, Calcutta
700 064\\
Saha Institute of Nuclear Physics, 1/AF Bidhan Nagar, Calcutta
700 064}
\date{\today}
\parindent=20pt
\maketitle

\begin{abstract}
The production of low mass dileptons and soft photons from thermalized
Quark Gluon Plasma (QGP) and hadronic matter in relativistic heavy ion 
collisions is evaluated. A boost invariant longitudinal and cylindrically
symmetric transverse expansion of the systems created in central collision of
lead nuclei at CERN SPS, BNL RHIC, and CERN LHC, and undergoing a first order
phase transition to hadronic matter is considered. A large production of
low mass ($M<$ 0.3 GeV) dileptons, and soft photons ($p_T<$ 0.4 GeV) is seen
to emanate from the bremsstrahlung of quarks and pions. We find an increase
by a factor of 2--4 in the low mass dilepton and soft photon
 yield as we move from SPS to RHIC
energies, and an increase by an order of magnitude as we move from
SPS to LHC energies. Most of the soft radiations are found to originate
from pion driven processes at SPS and RHIC energies, while at the LHC energies
the quark and the pion driven processes contribute by  a similar amount.
The study of the transverse mass distribution is seen to provide interesting
details of the evolution.
We also find a unique universal behaviour
for the ratio of $M^2$ weighted transverse mass distribution for $M=$ 0.1 GeV
to that for $M=$ 0.2 and 0.3 GeV, as a function of $M_T$, for SPS, RHIC, and
LHC energies, in the absence of transverse expansion of the system. A deviation
from this universal behaviour is seen as a clear indication of the flow.
\end{abstract}
\narrowtext

\section{INTRODUCTION}

Production, detection, and study of quark-gluon plasma (QGP) constitutes one of
the most important challenges of present day nuclear physics. 
There are plausible reasons to believe that this deconfined state of strongly
interacting matter may be produced in collisions involving heavy nuclei. This
expectation has led to a great deal of excitement, and a number of international
collaborative efforts are underway
to identify the signatures of QGP. Of these, 
lepton pairs ($e^+e^-$ or $\mu^+ \mu^-$) and photons are considered 
as one of the more 
reliable probes of this hot and dense phase since their
mean free path is quite large compared to typical nuclear size enabling
them to escape without any final state interaction.
Their abundance and spectral distributions are also a rapidly varying
function of the temperature  and thus they furnish most valuable information
about the nascent plasma. 

Spurred by these expectations a considerable  theoretical
effort has been devoted to the
study of large mass dileptons and high $p_T$ photons, which may have their
origin mostly in the early stages of the QGP.
A number of experiments
\footnote{ 
\noindent Was this the face that launched a thousand ships?\\
And burnt the topless towers of Ilium?\\
Sweet Helen, make me immortal with a kiss..\\
 -- Faustus, Christopher Marlowe (1564--1593)
}, viz.,
WA80, WA93, WA98, HELIOS, CERES, and NA38 experiments at
the CERN SPS, the PHENIX experiment at the BNL RHIC,
and the ALICE experiment at the  CERN LHC, are designed towards
measuring the electromagnetic radiations from relativistic heavy ion collisions.
 It should be remembered, though,
that photons and dileptons are emitted at every stage of the evolution of the
system, and they carry rather more precise
imprints of the circumstances of their
`birth'. We shall see that the intense glow of soft photons and low mass
dileptons can provide reliable and useful information about the later
stages of the interacting system.

We concentrate in particular on soft photons and low-mass dileptons 
produced by bremsstrahlung processes whose
energies are low enough to enable us to use the  so-called soft-photon
approximation \cite{ruckl}. This brings in a unique advantage as, 
but for a `known' phase
space factor (see later) arising due to the finite mass of the dileptons, the
basic cross-sections for the two processes become identical. The dileptons
of different invariant masses, however, are affected differently by 
the transverse flow,
which should be substantial towards the last moments of the interacting system.
 A comparison of the yield of dileptons of different masses and photons could 
then provide us with a valuable information about the flow. This
identity of the cross-section for 
the basic process is not available, for example, between 
single photons originating from Compton and annihilation processes 
(in the QGP), and nuclear reactions of the type, say, 
$\pi\rho\,\rightarrow\,\pi \gamma$  in the hadronic matter or
quark or pion annihilation processes for the dileptons.

It is also important to understand this contribution in quantitative detail
as it is quite likely that the lowering of the mass of $\rho$ mesons,
due to high baryonic densities reached in such collisions, for
example,  will populate the mass region well below the $m_\rho$
in case of dileptons.  Thus for example, Li et al
\cite{li} postulate that the mass of the `primordial' $\rho$ mesons may have
 dropped to 370 MeV in S+Au collisions at the CERN SPS studied by the 
CERES group \cite{ceres}. The drop is likely to be even larger for the Pb+Pb
collsion at the SPS energies, in  their treatment \cite{ko}.
This increase in the net baryonic density is unlikely to be achieved 
at the RHIC or the LHC energies because of the considerable increase in the
transparency.

On the other hand, there are reasons to believe that the pion-form factor
 $F_\pi(M)$, as
well as the decay width ($\Gamma_\rho$) of the $\rho$ meson, e.g.,  
may depend on the
temperature, either because of chiral symmetry restoration, or collision
broadening, or both. Thus, in the simplest approximation for the chiral
perturbation theory, modifications of $F_\pi$ and $\Gamma(\rho)$ are given
by \cite{gaber}
\begin{eqnarray}
F_\pi(M,T)&=&F_\pi(M,0)\left( 1-\frac{T^2}{8F_\pi^2(M,0)}\right)~,\nonumber\\
\Gamma_\rho(T)&=&\Gamma_\rho(0)/ \left(1-\frac{T^2}{4F_\pi^2(M,0)}\right),
\end{eqnarray}
with the mass of the $\rho$ meson remaining essentially unchanged (see e.g.,
Ref. \cite{rob} for more recent developments). 
This corresponds to a
decrease in $F_{\pi}$ by about 30\% and an increase in $\Gamma_{\rho}$ by
a factor of 3 at T=160 MeV, thus affecting the production of dileptons
from annihilation of pions beyond $M\approx$ 400 MeV or so.
The broadening of the $\Gamma(\rho)$ 
due to collisions has been estimated by Haglin
\cite{kevin2} as
\begin{equation}
\Gamma_\rho(T)=\Gamma_\rho(0)+\left(a+b T +c T^2\right),
\end{equation}
where $a= 0.50$ GeV, $b=-7.16$, and $c=30.16$ GeV$^{-1}$, and corresponds to an 
increase of about 80\% at $T=$ 160 MeV, whose effect will be limited to
dileptons from pion annihilation beyond $M\approx$ 500 MeV. 

Recall that in the early days of the investigations of QGP it was 
often suggested
that thermal dileptons having an invariant mass less than $2 m_\pi$ could
originate only from the annihilation of quarks and anti-quarks which were
assumed to be essentially massless.
 Very soon it was realized (see Ref.\cite{helmut,kevin1} and references
therein) that there
could be substantial production of dileptons having lower invariant
masses from the bremsstrahlung of pions as well as quarks.
A better understanding \cite{braaten,weldon} of the dynamics of hot QCD has 
endowed quarks with a thermal mass of a few hundred MeV. 
If we believe \cite{kamp} that
the invariant mass of the dileptons will have $M \geq 2 m_{\mathrm {th}}$,
if they originate from the quark annihilation, then the mass window below a
few hundred MeV is  populated primarily by dileptons from 
bremsstrahlung processes at the colliders. Of course, there would be a
background from Dalitz decays of $\eta$ and $\pi^0$ mesons, which will have to
be eliminated before the glow of the soft dileptons becomes visible.
Similar considerations will apply to soft photons originating from the
bremsstrahlung of pions.

   The QGP likely to be produced in relativistic heavy ion collisions will
have enormous internal pressure, and will expand rapidly \cite{vesa}.
 If the life time of the interacting system is large compared to
 $\sim R_T/c_s$, where
$R_T$ is the transverse size of the system and $c_s$ is the speed 
of sound, the consequences of the transverse expansion will become very evident.
The last stages of the interacting system are also likely to be repository
of the details of the flow, and thus the soft photons and low mass
dileptons which derive their maximum contributions from this stage will
also carry unique information about the flow.

   We organise our paper as follows. In Section II we briefly recall
the formulation for the bremsstrahlung production of soft photons and low
mass dileptons.
Section III describes the results and discussions  related to various
approximations used in the work are given in Section IV.  
Finally we give a brief
Summary. With this we also conclude our study of soft electromagnetic
radiations \cite{dipali,pradip}
initiated earlier.

\section{FORMULATION}

\subsection{Soft Photon Approximation}

The production of low mass dileptons and soft photons is most conveniently
evaluated within a soft-photon approximation. In so far as this approximation
remains valid, it provides for an easy manipulation of the strong interaction
part of the scattering and enables us to test the sensitivity of the
results to such details. 

Thus the first question which comes to mind is, how
reliable is the soft-photon approximation? The existing treatments for the
bremsstrahlung production of low mass dileptons were critically examined by
Lichard, recently \cite{peter}. We shall, as in our earlier works
 \cite{dipali,pradip}, use the correct numerical factor of 
($\alpha/3\pi M^2$) in Eq.(8) (see later)
and also use the virtual photon current, as suggested by Lichard.

A reasonably accurate check on the soft photon approximation for
the quark driven processes can be obtained from a comparison with the
rate for a zero-momentum soft dilepton production \cite{eric} in a QGP
evaluated by using the resummation technique of 
Braaten and Pisarski \cite{braaten}.  This was done recently \cite{kevin1},
with interesting results. To quote, it was found that for dileptons having
masses $M\leq$ 0.1 GeV, the soft photon approximation gives results
which are very close to the findings of QCD perturbation theory.
The soft photon approximation was found to lead to results which were
smaller by a factor of about 1--4 for $0.1\leq M \leq 0.3$ GeV.
This has its origin in the fact that the QCD perturbation
theory includes the annihilation process, which
contributes substantially at larger masses (see figs.1a--c, later).
 This comparison  provides two important insights;
viz., scattering with virtual bremsstrahlung
(rather than annihilation or Compton-like processes) accounts for most of the 
low-mass QGP-driven pairs and  the soft-photon approximation as applied to
quark processes is fairly reasonable. We must add that we
shall depict lowest order annihilation process $q\bar{q}\rightarrow e^+e^-$
clearly and separately. 

In a recent study, Eggers et al. \cite{hans} have evaluated the bremsstrahlung
production of dileptons from pion driven processes, without using the
soft-photon approximation in a One Boson Exchange model for the 
interaction of pions. One of the many interesting observations in that work
is that the use of the soft-photon approximation in terms of the invariant
mass of the dileptons can overestimate the contribution of the bremsstrahlung
processes. We use the soft-photon approximation in terms of the four-
momenta of the real or the virtual photons, and thus we feel that we are
relatively safe from this criticism. Thus we insist that both $M$ and 
$q$ remain reasonbly small. We have also limited ourselves to
invariant masses upto 300 MeV. 
Still it is worthwhile to recall that even though the individual contributions
from different reactons involving pions to the basic cross-section
 ($d\sigma/dM$)  are off by differing amounts as compared to the predictions
of the soft-photon approximation, the  rates are overestimated by atmost
 a factor of 1.5--2 for $M\leq$ 0.3 GeV, provided we follow the suggestions
of Lichard \cite{peter}, as we have.

In view of the above discussion, we believe that the soft-photon approximation
as employed by us is reliable to within a factor of 2.
Still it should be certainly worthwhile to have the 
results of this treatment \cite{hans} for the transverse mass distribution to
settle this issue, clearly.

\subsection{Low Mass Dileptons}

The mechanism for the production of soft  virtual photons from
bremsstrahlung within a soft photon approximation  
has been discussed by a number 
of authors ( see Ref.\cite{dipali} and references therein) 
in great detail and thus we shall only briefly recall the
formulation in order to fix the notation.

 The invariant cross-section for the scattering and at the same time
production of a soft photon of four momentum $q^\mu=(q^0,\vec{q})=(E,\vec{q})$ 
is given by
\begin{equation}
q_0 \frac{d^4 \sigma^{\gamma}}{d^3q dx} = \frac{\alpha}{4 \pi^2}
\left \{ \sum_{{\mathrm {pol}} \lambda} J \cdot \epsilon_{\lambda}~~ 
J \cdot \epsilon_{\lambda} \right \}\frac{d\sigma}{dx}
\end{equation}
where $d\sigma/dx$  is the strong interaction cross-section for the
reaction $ab\,\rightarrow\, cd$, $\epsilon_\lambda$ is the polarization of the
emitted photon, and
 $J^\mu$ is the {\em virtual photon current}\cite{peter} given by
\begin{eqnarray}
J^\mu&  =& -Q_a \frac{2p_a^{\mu}-q^{\mu}}{2p_a \cdot q-M^2}-
Q_b \frac{2p_b^{\mu}-q^{\mu}}
{2p_b \cdot q-M^2}\nonumber\\
& &+Q_c \frac{2p_c^{\mu}+q^{\mu}}{2p_c \cdot q+M^2}+
Q_d \frac{2p_d^{\mu}+q^{\mu}}{2p_d \cdot q+M^2}.
\end{eqnarray}
In the above equation the $Q$'s and $p$'s  represent the charges (in units of
proton charge) and the particle four momenta, respectively. 
The cross-section for the production of dilepton is then obtained as
\begin{equation}
E_+E_-\, \frac{d^6 \sigma^{e^+e^-}}{d^3p_+d^3p_-} = \frac{\alpha}{3\pi^2}
\frac{1}{q^2}\, q_0\frac{d^3\sigma^{\gamma}}{d^3q}
\end{equation}

Now the invariant cross-
section for dilepton pair production can be written as \cite{dipali}
\begin{eqnarray}
E_+E_-\, \frac{d\sigma_{ab \rightarrow cd}^{e^+e^-}}{d^3p_+d^3p_-}=
\,& & \frac{\alpha^2}{12 \pi^4 M^2}   \int   
 |{\epsilon \cdot J}|_{ab \rightarrow cd}^2
\,\frac{d\sigma_{ab \rightarrow cd}}{dt}\nonumber\\
&\times & \delta(q^2-M^2)\, dM^2\,\nonumber\\
& \times& \delta^4 \left(q-(p_++p_-)\right) \,d^4q \,dt.
\end{eqnarray}
The rate of production of dileptons at temperature $T$
can then be written as
\begin{eqnarray}
E\frac{dN}{d^4xdM^2d^3q}=\frac{T^6g_{ab}}{16\pi^4} 
\int_{z_{\mathrm {min}}}^{\infty} \,dz & &
\frac{\lambda(z^2T^2,m_a^2,m_b^2)}{T^4}\nonumber\\
&\times &
\Phi(s,s_2,m_a^2,m_b^2)\,\nonumber\\
&\times & K_1(z)\,
E\frac{d\sigma_{ab}^{e^+e^-}}{dM^2d^3q},
\end{eqnarray}
where
the cross-section for the process $ab \rightarrow cd \,e^+e^-$
is given by
\begin{equation}
E\frac{d\sigma_{ab \rightarrow cd}^{e^+e^-}}{dM^2d^3q} =
 \frac{\alpha^2}{12 \pi^3 M^2} \,\frac{\widehat{\sigma}(s)}{E^2},
\end{equation}
with
\begin{equation}
\widehat{\sigma}(s)=\int_{- \lambda(s,m_a^2,m_b^2)/s}^0 \,dt\,
\frac{d\sigma_{ab \rightarrow cd}}{dt}\,
\left(q_0^2\left|\epsilon\cdot J\right|^2_{ab \rightarrow cd}\right),
\end{equation}
and
\begin{equation}
\Phi(s,s_2,m_a^2,m_b^2)=\frac{\lambda^{1/2}(s_2,m_a^2,m_b^2)}
                         {\lambda^{1/2}(s,m_a^2,m_b^2)}\,\frac{s}{s_2},
\end{equation}
$s_2=s+M^2-2\sqrt{s}q_0$, and $\lambda(x,y,z)=x^2-2(y+z)x+(y-z)^2$.
The expression for the average of the electromagnetic factor 
over the solid angle, can be found in Ref.\cite{dipali}. 
The value of $z_{\mathrm {min}}$ is obtained from 
$\lambda(s_2,m_a^2,m_b^2) = 0$. Note that the right hand side of Eq.(8)
varies as $1/M^4$ for ${\mathbf q}=0$.

     The strong interaction differential cross-section
$d\sigma_{qq}/dt$ and $d\sigma_{qg}/dt$ for scattering of
quarks and gluons are obtained from semi-phenomenological expressions
used earlier by several authors for this purpose
 \cite{kevin1,dipali,pradip,daniel}. For hot hadronic matter, we
have included the leading reactions: $\pi^+ \pi^- \rightarrow \pi^0
\pi^0$,  $\pi^+ \pi^- \rightarrow \pi^+ \pi^-$,  $\pi^+ \pi^0 \rightarrow
\pi^+ \pi^0$,  and  $\pi^- \pi^0 \rightarrow \pi^- \pi^0$ and evaluated
the strong scattering cross-section from an effective Lagrangian
incorporating $\sigma,$  $\rho,$ and $f$ meson exchange 
\cite{kevin1,dipali,pradip}.

 The corresponding expressions for the contribution of annihilation
processes $q \bar q \rightarrow e^+e^-$ and $\pi^+ \pi^- \rightarrow
e^+e^-$ are given \cite{kkmm} by,

\begin{eqnarray}
E\frac{dN}{d^4x dM^2 d^3q}& =& \frac{\sigma_a(M)}{4 (2\pi)^5} M^2 e^{-E/T}
\left[1-\frac{4m_a^2}{M^2}\right],\nonumber\\
\sigma_{a}(M) &= &F_{a} {\bar \sigma(M)}, \nonumber\\
{\bar \sigma(M)}& = &\frac{4 \pi \alpha^2}{3 M^2}
\left[1+\frac{2m_e^2}{M^2}\right] \left[1-\frac{4m_e^2}{M^2}\right]^{1/2},
\end{eqnarray}
where $F_q=20/3$ for a QGP consisting of $u$ and $d$ quarks, and
gluons, and $F_\pi$ is the pion form factor.

\subsection{ Soft Photons}

Now we consider soft photon emission through the 
bremsstrahlung process, $ab \rightarrow
cd \gamma$.
The invariant cross-section for the above process is obtained from
eq. (4) with $J^{\mu}$ replaced by 
\begin{equation}
J^\mu = -Q_a \frac{p_a^{\mu}}{p_a \cdot q}-Q_b \frac{p_b^{\mu}}
{p_b \cdot q}+Q_c \frac{p_c^{\mu}}{p_c \cdot q}+
Q_d \frac{p_d^{\mu}}{p_d \cdot q},
\end{equation}
which is appropriate for the emission of real photons.
The rate of production of photons at temperature $T$
can then be written as
\begin{eqnarray}
E\frac{dN}{d^4xd^3q}=\frac{T^6g_{ab}}{16\pi^4}
 & &\int_{z_{\mathrm {min}}}^{\infty} \,dz
\frac{\lambda(z^2T^2,m_a^2,m_b^2)}{T^4}\nonumber\\
&\times &
\Phi(s,s_2,m_a^2,m_b^2)\, K_1(z)\,
E\frac{d\sigma_{ab}^{\gamma}}{d^3q},
\end{eqnarray}
where
\begin{equation}
E\frac{d\sigma_{ab}^{\gamma}}{d^3q} = \frac{\alpha}{4\pi^2}\, 
\frac{{\widehat
\sigma}(s)}{E^2},
\end{equation}
with ${\widehat\sigma}(s)$ defined as before (Eq.(9)) with $J^\mu$ 
replaced by real photon
current Eq.(12).
Even at the risk of repetition, we would like to add that {\it if we 
put $M = $ 0 in the phase-space factor $\Phi_2$ and use the real photon
current in Eq.(7) and Eq.(9)}, we shall have
\begin{equation}
E \frac{dN_{e^+e^-}}{d^4xdM^2d^3q} \equiv \frac{\alpha}{3 {\pi} M^2}\,\,
E \frac{dN_{\gamma}}{d^4xd^3q},
\end{equation}
which also remains true in limit $M \rightarrow $0.
Thus a comparison of the expression Eq.(13) with Eq.(7) 
immediately shows that one
may use the results for photons and dileptons (with different masses) 
with advantage to get information about, say,  the evolution of the system.

The annihilation and the Compton processes $q \bar q \rightarrow \gamma g$
and $q(\bar q)g \rightarrow q(\bar q)g \gamma$ have already been studied
in great detail by a number of authors \cite{Kapusta}. We only mention 
the result for a comparison:
\begin{equation}
E\frac{dN_{\gamma}^{C+ann}}{d^4xd^3q} = \frac{5}{9}\, \frac{\alpha \alpha_s}
{2\pi^2}\, T^2\, e^{-E/T}\, \ln \left( \frac{2.912ET}{6m_q^2}+1 \right)
\end{equation}
where $m_q = \sqrt{ 2\pi \alpha_s/3}\,T$ is the thermal mass of the
quarks.
In the hadronic sector we consider the processes
$\pi \rho \rightarrow a_1 \rightarrow \pi \gamma$, $\pi \rho
\rightarrow \pi \gamma$ 
 for which rates
have been evaluated and parametrized in a convenient form 
\cite{Kapusta,Xiong,Nadeu}.   
\begin{equation}
E\frac{dN_{\pi \rho \rightarrow a_1 \rightarrow \pi \gamma}}{d^4xd^3q} =
 2.4T^{2.15}\,\exp\left[-1/(1.35TE)^{0.77}-E/T\right]\\,
\end{equation}
\begin{equation}
E\frac{dN_{\pi \rho  \rightarrow \pi \gamma}}{d^4xd^3q} =
 T^{2.4}\,\exp\left[-1/(2TE)^{3/4}-E/T\right]\\.
\end{equation}
The decay  $\omega \rightarrow \pi \gamma$ during the life-time of the
interacting system is obtained from,
\begin{eqnarray}
E\frac{dN_{\omega  \rightarrow \pi \gamma}}{d^4xd^3q} =
\frac{3m_{\omega}\Gamma_{\omega \rightarrow \pi \gamma}}{16\pi^3E_0E}
& &\int_{E_{min}}^{\infty}\, 
 dE_{\omega}\,E_{\omega}f_{B E}(E_{\omega})
\nonumber\\
&\times &\left[1+f_{B E}(E_{\omega}-E)\right]
\end{eqnarray}

Here $E_{min} = m_{\omega}(E^2 + E^2_0)/2EE_0$ and $E_0$ is the photon
energy in the rest frame of the $\omega$ meson. Recall that for low energy
photons the reactions $\pi\pi\,\rightarrow\,\rho\gamma$ and
the bremsstrahlung process $\pi\pi\,\rightarrow\, \pi\pi\gamma$ are equivalent,
and including both of them would amount to a double
counting \cite{pradip,redlich}.

\subsection{ Initial Conditions}

 We have considered central collisions of lead nuclei at CERN SPS, BNL RHIC,
and CERN LHC energies. We assume that the collision leads to a thermalized
and chemically equilibrated quark gluon plasma at an initial time
 $\tau_i=$ 1 fm/$c$ and initial temperature $T_i$.
 Further assuming an isentropic
expansion, one may relate \cite{hk} the initial conditions to the 
multiplicity density ($dN/dy$);
\begin{equation}
T_i^3 \tau_i = \frac{2 \pi^4}{45 \zeta(3) \pi R_T^2 4a_k}\, \frac{dN}{dy},
\end{equation}
where $R_T$ is the transverse radius of the lead nucleus and 
$a_k = 37\pi^2/90$ for a system consisting
of massless u and d quarks, and gluons. The evolution
of the system is obtained from a boost-invariant longitudinal
expansion and cylindrically symmetric transverse expansion \cite{vesa}.
We further assume a first order phase transition to a hadronic matter 
consisting of  $\pi$, $\rho$, $\omega$, and $\eta$ mesons,
($a_k \approx 4.6\pi^2/90$), at $T=$ 160 MeV \cite{our}.
After all the quark matter has adiabatically converted to hadronic matter,
the system enters a hadronic phase and undergoes a freeze-out at $T=$ 140 MeV.

The particle rapidity density is taken as \cite{kms} 624, 1735, and 5624
respectively. One may obtain much larger initial temperatures for the
same multiplicity densities by assuming more rapid thermalization of
the plasma. An upper limit for this is obtained by taking 
$\tau_{i} \simeq 1/3 T_{i}$. We shall argue later, that the multiple
scattering effects in the early dense QGP will, however, suppress the
soft radiations considerably, and thus  the choice of
$\tau_i$ =1 fm/$c$ should provide an interesting trade-off between
these competing effects.

\subsection{ Space-time Integration}

The dilepton transverse mass yield is then obtained by convoluting
the rates for their emission from QGP and hadronic matter with the space- 
time history of the system:
\begin{eqnarray}
\frac{dN}{dM^2d^2M_Tdy}=\int & &\,\tau \,d\tau\, r\, dr\, d\phi\, d\eta
\left[f_Q\,E\frac{dN^q}{d^4x dM^2 d^3q}\right.\nonumber\\
&+&\left.(1-f_Q)\,E\frac{dN^{\pi}}
{d^4x dM^2 d^3q}\right]~,
\end{eqnarray}
where $f_Q(r, \tau)$ gives the fraction of the quark matter
in the system.

   Similarly the photon spectrum is obtained by convoluting the rates for the
emission of photons from QGP and the hadronic matter with the space time
history of the system;  
\begin{eqnarray}
\frac{dN}{d^2q_Tdy}=\int & &\,\tau \,d\tau \,r \,dr\, d\phi \,d\eta
\left[f_Q\, E\frac{dN^q}{d^4x d^3q}\right.\nonumber\\
&+&\left.(1-f_Q)\,
 E\frac{dN^{\pi}}{d^4x d^3q}\right].
\end{eqnarray}

\section{RESULTS}

\subsection{Low mass Dileptons }

In order to ascertain the relative importance of the contributions of the
 quark bremsstrahlung, 
pionic bremsstrahlung, quark annihilation, and pionic annihilation processes
to low mass dileptons we show the rates for different values of $M$
at $T=$ 160 MeV. All the results for the quark annihilation processes are
obtained by taking $m_q=$ 5 MeV. If we adopt the view that 
$m_q=m_{\mathrm {th}}$,
as indeed, we have taken while evaluating the bremsstrahlung contributions,
 then the quark annihilation contribution will be absent
\cite{kamp} in this mass range.
In any case, we see that the quark driven bremsstrahlung processes
outshine the pion driven bremsstrahlung contribution (fig.1a--c).

\begin{figure}
\psfig{figure=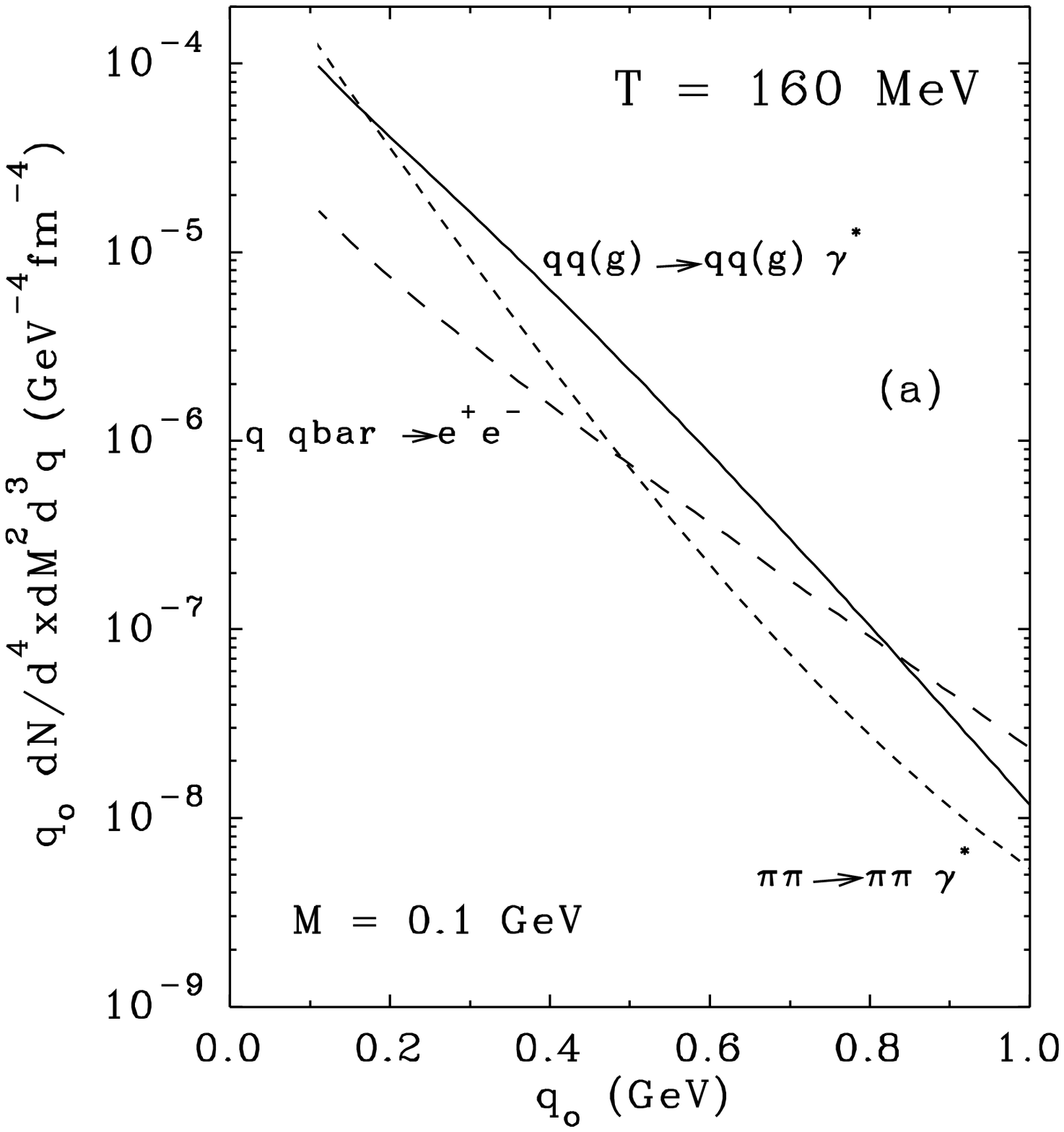,height=2.25in,width=3.25in}
\vskip 0.2cm
\psfig{figure=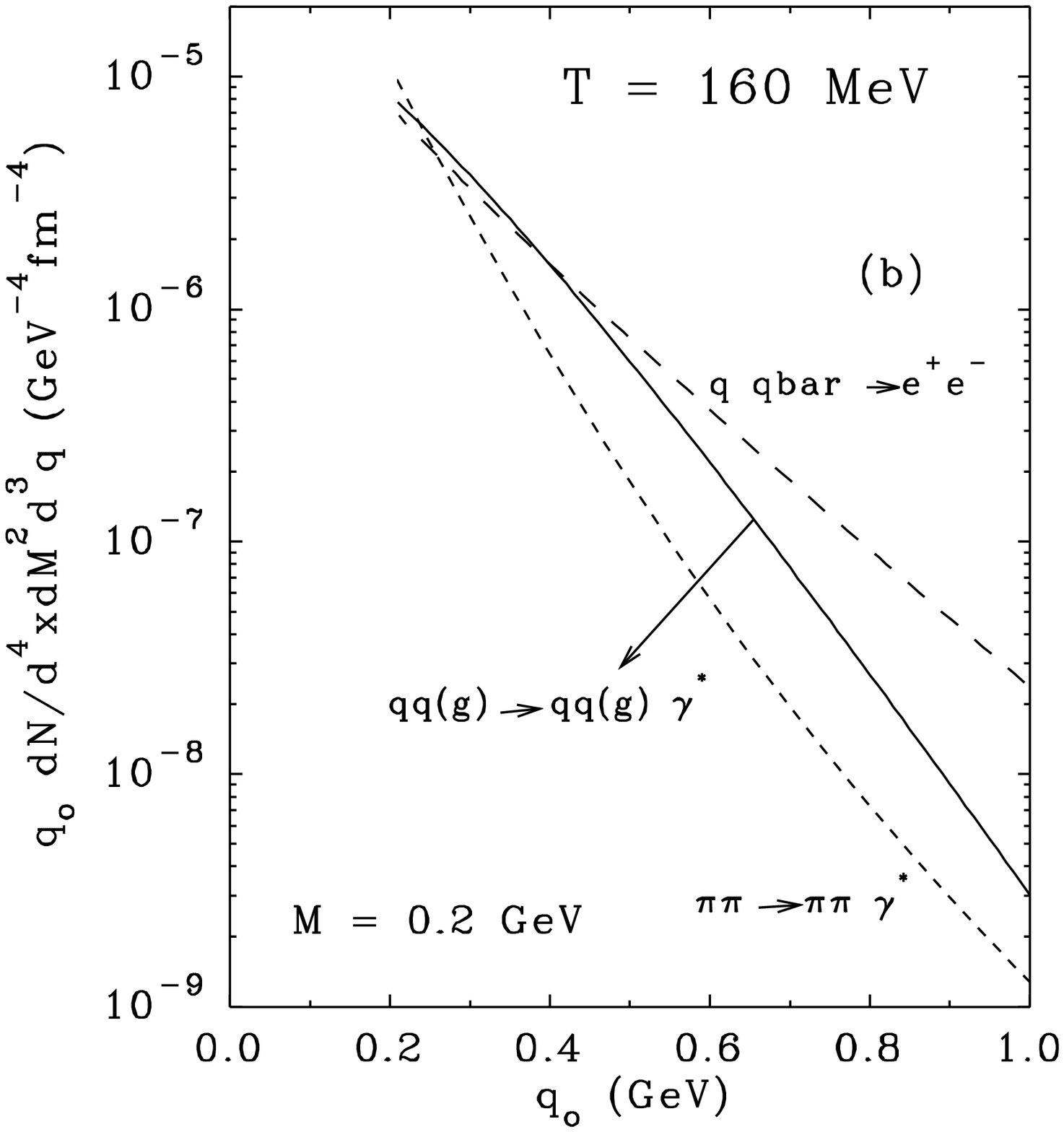,height=2.25in,width=3.25in}
\vskip 0.2cm
\psfig{figure=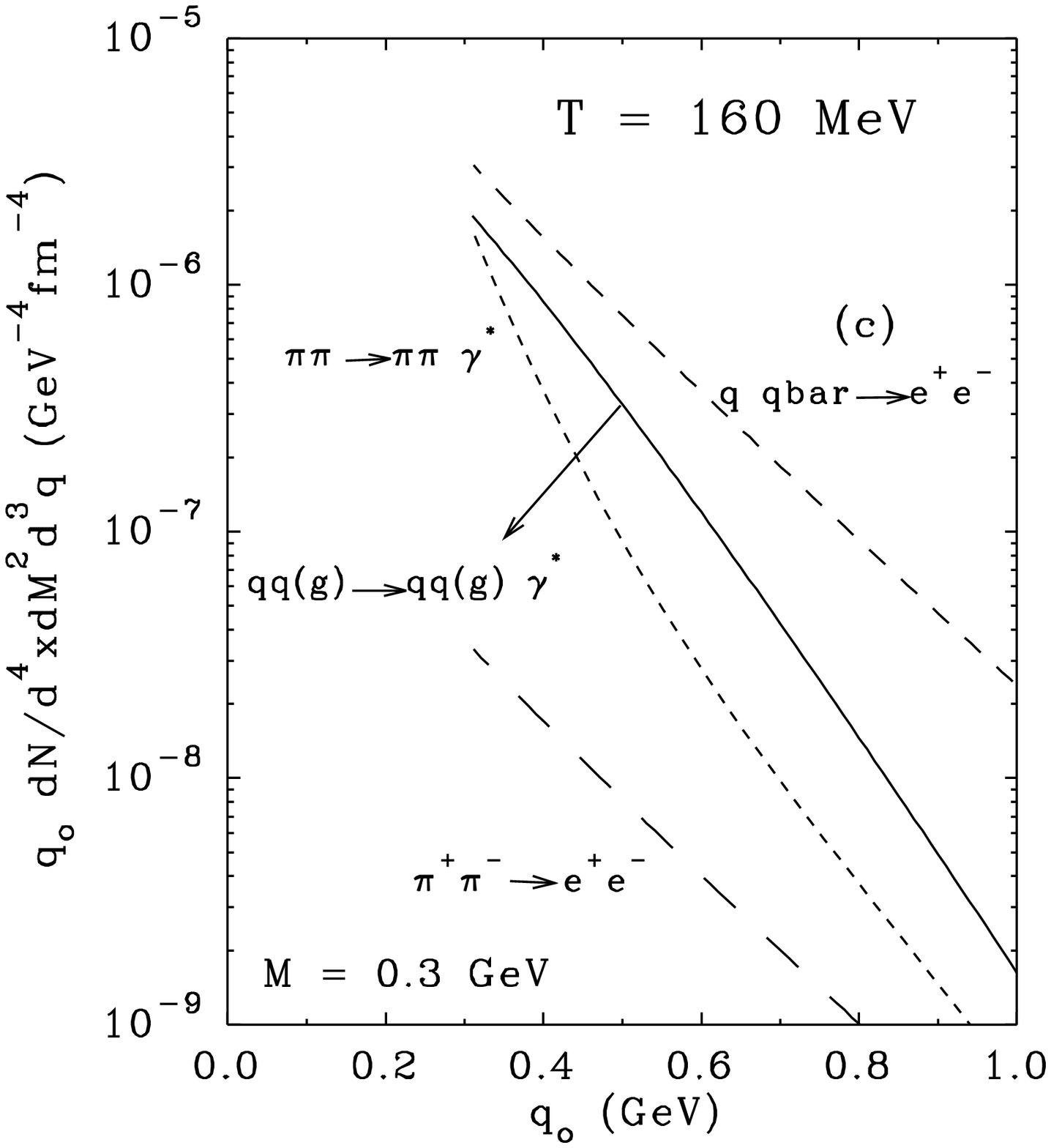,height=2.25in,width=3.25in}
\vskip 0.4cm
\caption{
 The production rate of low mass dielectrons from quark
and pion bremsstrahlung at $T = $ 160 MeV.
In addition, the contribution of quark annihilation process is given
for a comparison. These results are shown for (a) $M =$ 0.1 GeV, (b)
 $M =$ 0.2 GeV, and (c) $M =$ 0.3 GeV respectively.}
\end{figure}

The results for the transverse mass distribution for the low mass dileptons
at SPS energies are given in fig.2a--c.
We now see that the pion driven processes dominate the yield at all masses
as the 4- volume occupied by the hadronic matter is much larger.
This is also evident from the invariant mass distribution (fig.2d).
Considering that the pion annihilation threshold limits the mass to $M>2m_\pi$
and even the quark annihilation may contribute only to masses larger than this,
we do find an intense glow of low mass dileptons, once the background from
the Dalitz decays of $\pi^0$ and $\eta$ mesons is removed. The recent
experience with the CERES experiment \cite{ceres} has shown 
that this could be possible to some extent. Recall also that the CERES
data for the S+Au system \cite{ceres} shows a contribution from the
bremsstrahlung processes \cite{ssg}. It will be interesting to find a
confirmation of these early observations from the results for the Pb+Pb
system as well.

\begin{figure}
\psfig{figure=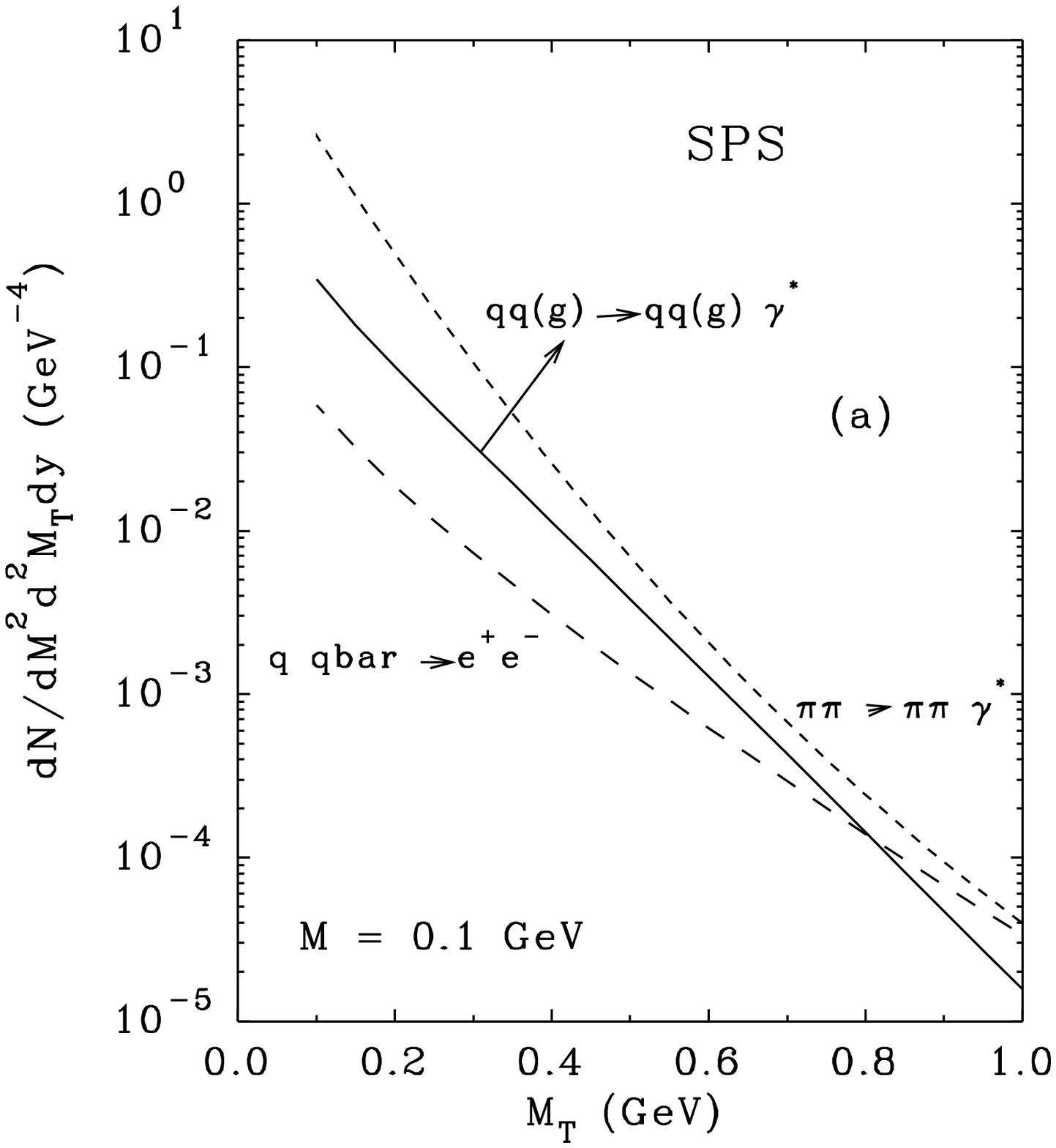,height=2.25in,width=3.25in}
\vskip 0.2cm
\psfig{figure=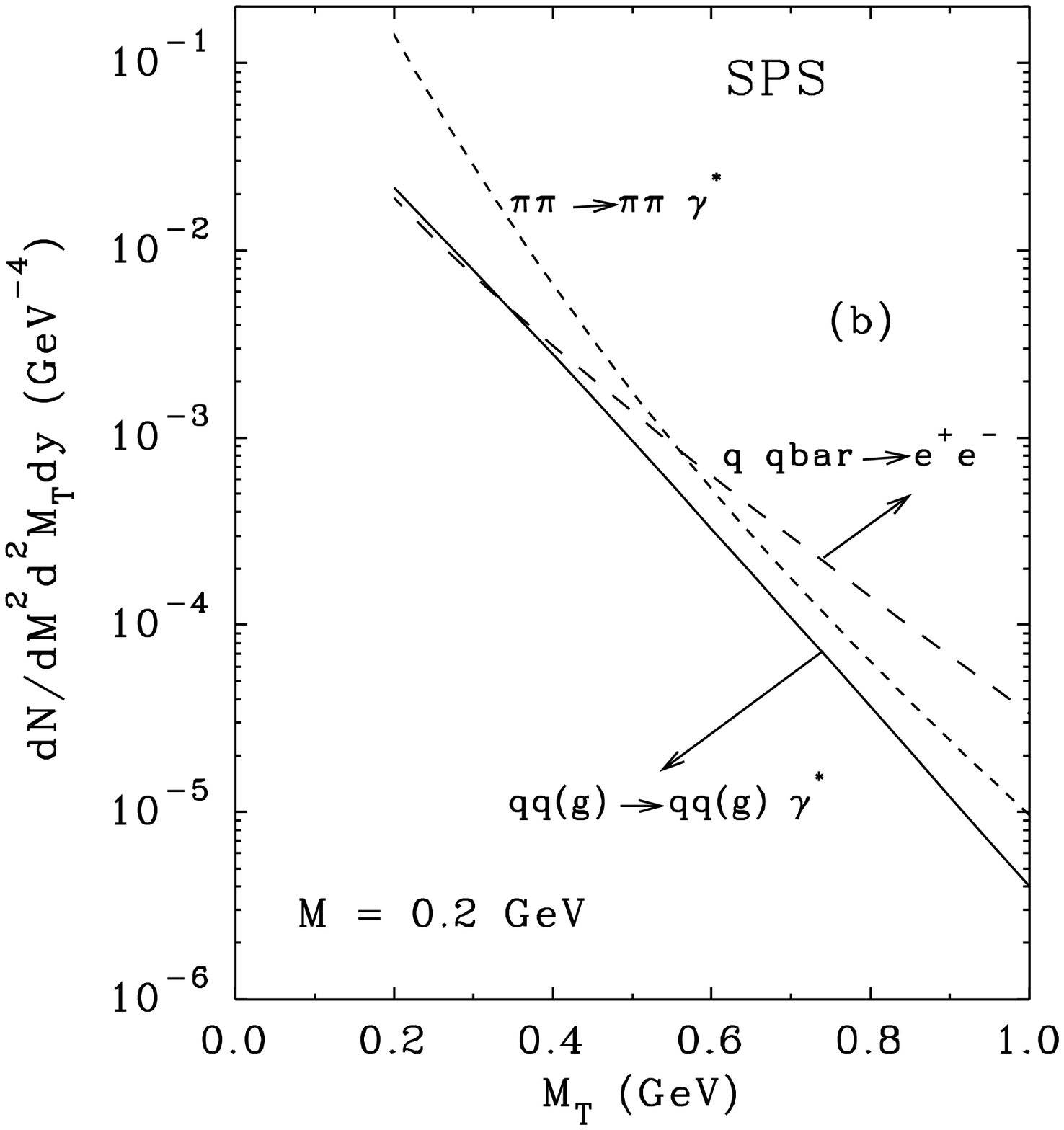,height=2.25in,width=3.25in}
\vskip 0.2cm
\psfig{figure=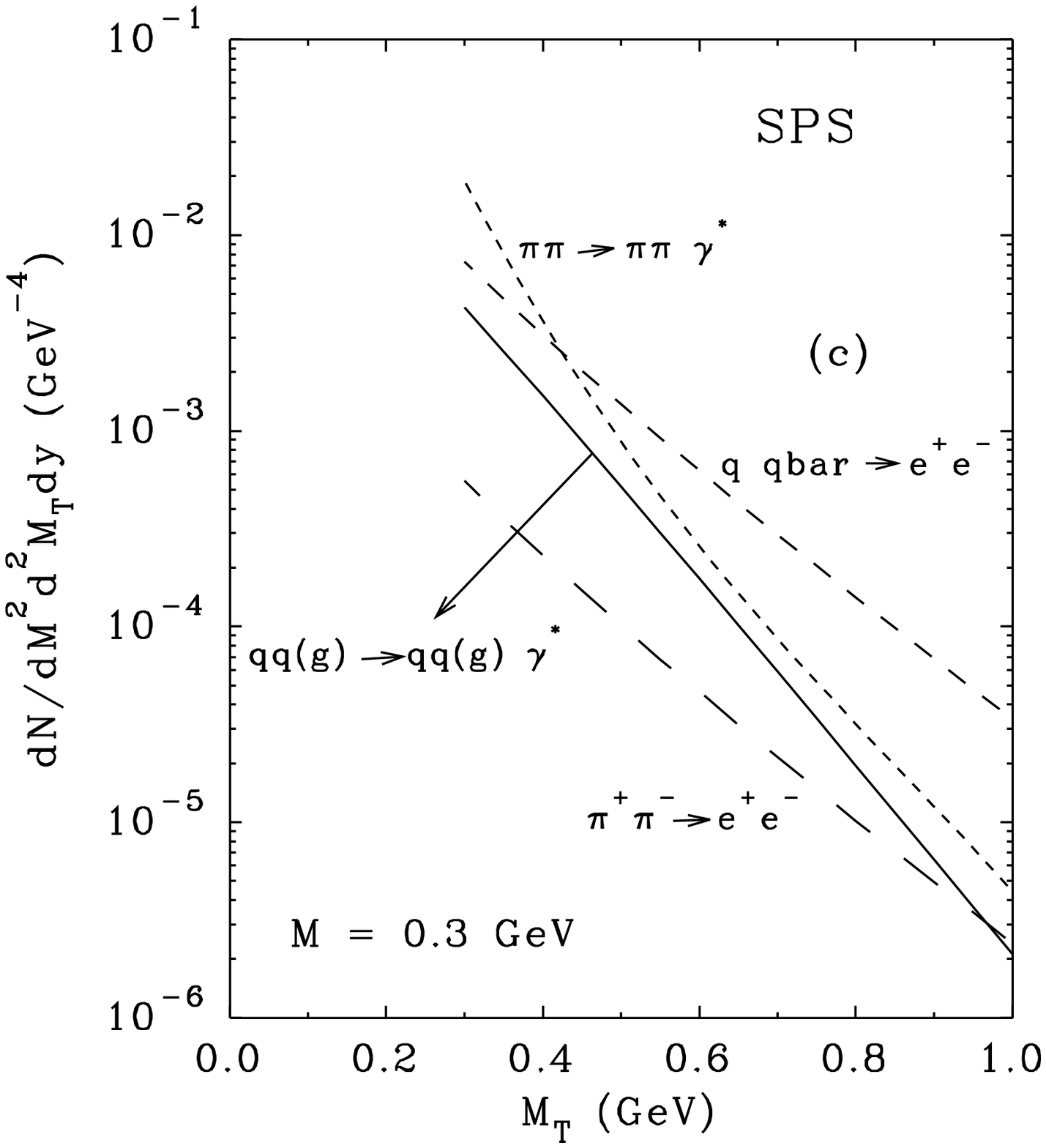,height=2.25in,width=3.25in}
\vskip 0.2cm
\psfig{figure=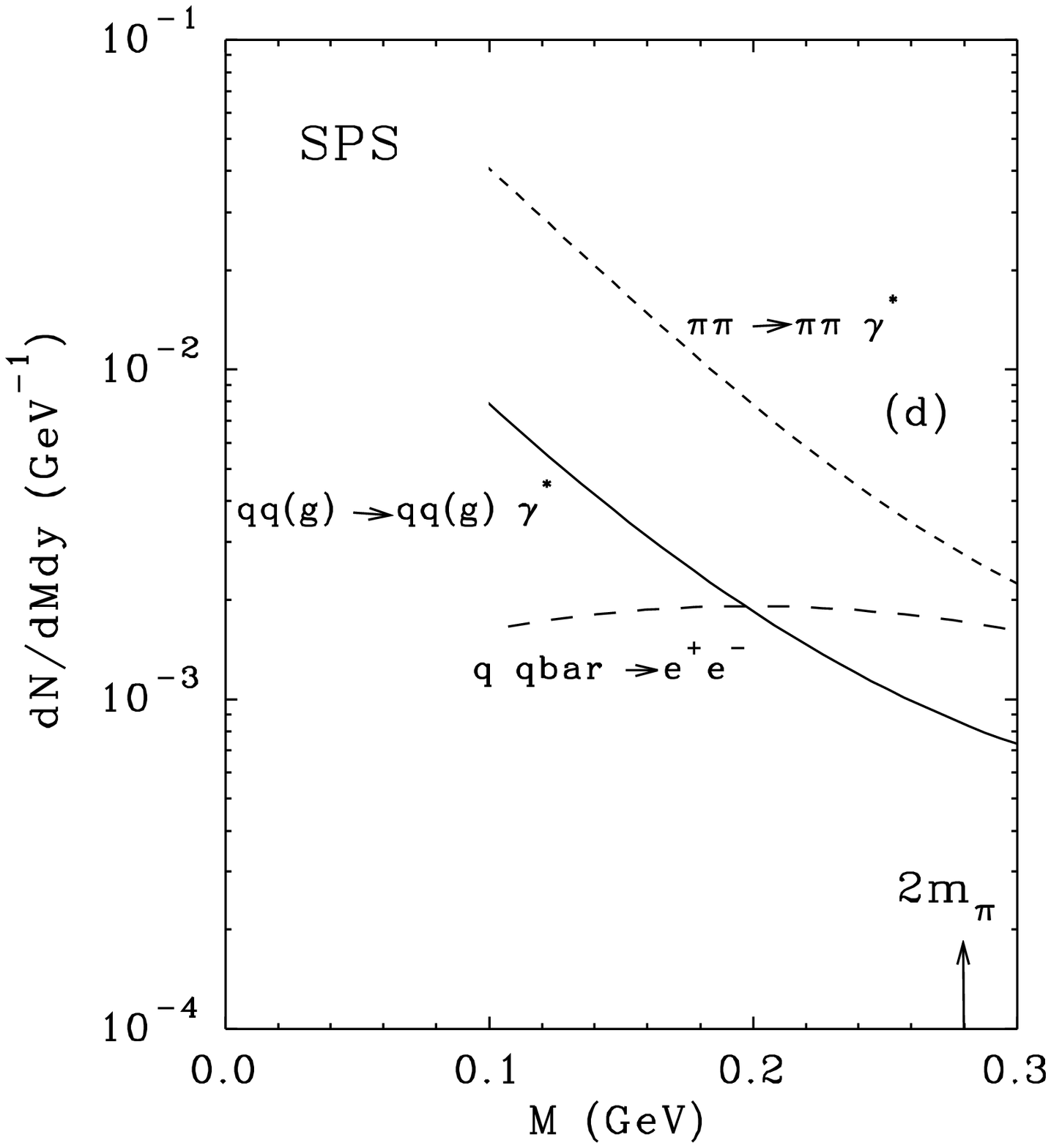,height=2.25in,width=3.25in}
\vskip 0.4cm
\caption{(a--d): The transverse mass distribution of low mass dielectrons
at SPS energies including bremsstrahlung process and annihilation 
process in the quark matter and the hadronic matter. We give the results
for invariant mass M equal to
 0.1 GeV (a), 0.2 GeV (b), and  0.3 GeV (c) 
respectively. The invariant mass distribution of low mass
dielectrons are also shown (d).}
\end{figure}

The transverse mass distribution at RHIC energies (fig.3a--c) reveals another
interesting aspect. The transverse mass distribution at lower $M_T$ is
dominated by the pion contribution. However at larger $M_T$, the contributions
of the quark driven and pion driven processes are similar. This is a 
reflection of the larger temperature in the quark phase, and a larger effect
of the transverse flow during the hadronic phase. If we look only at the 
invariant mass distribution (fig.3d), this interesting aspect does not show up.

\begin{figure}
\psfig{figure=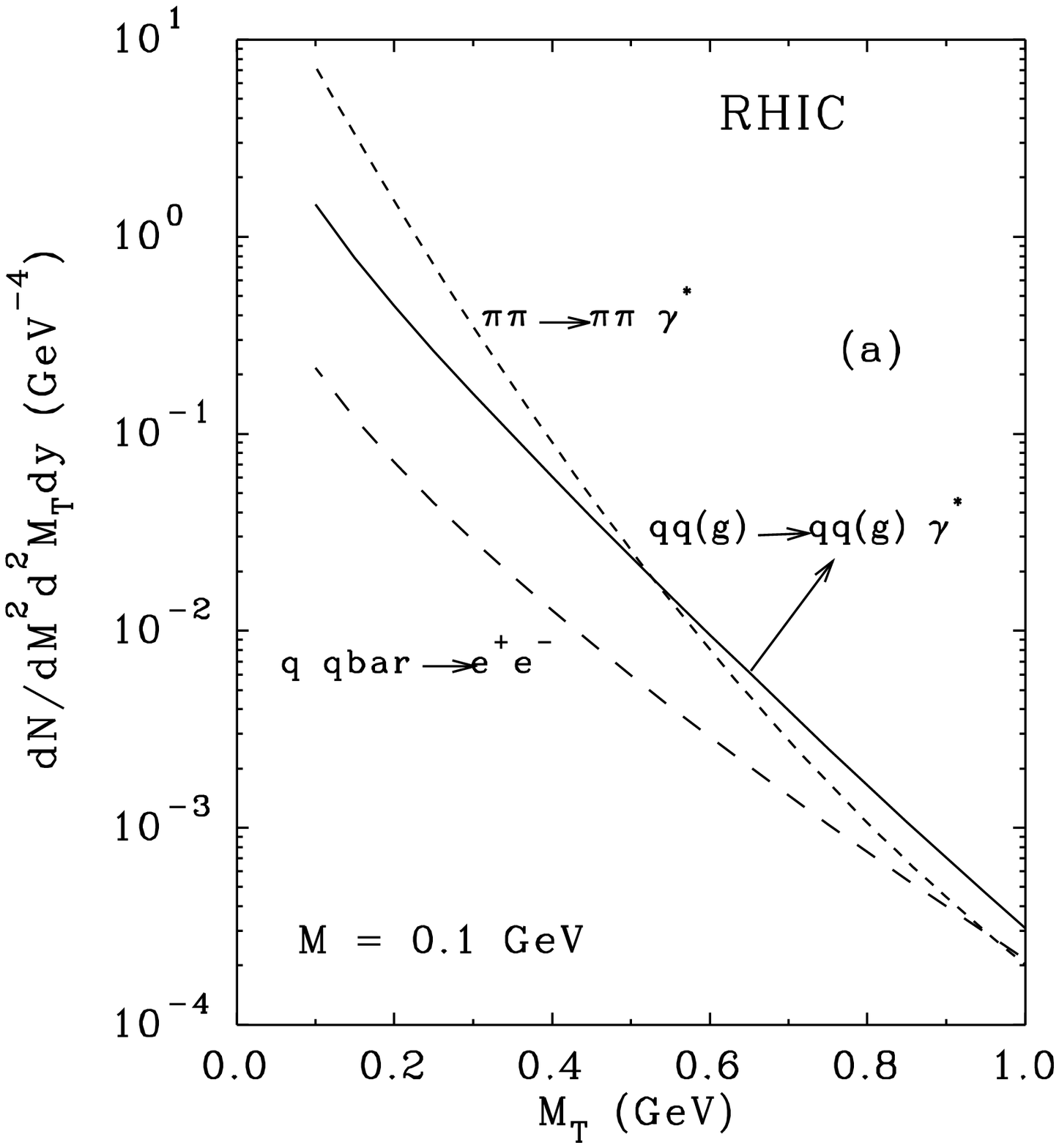,height=2.25in,width=3.25in}
\vskip 0.2cm
\psfig{figure=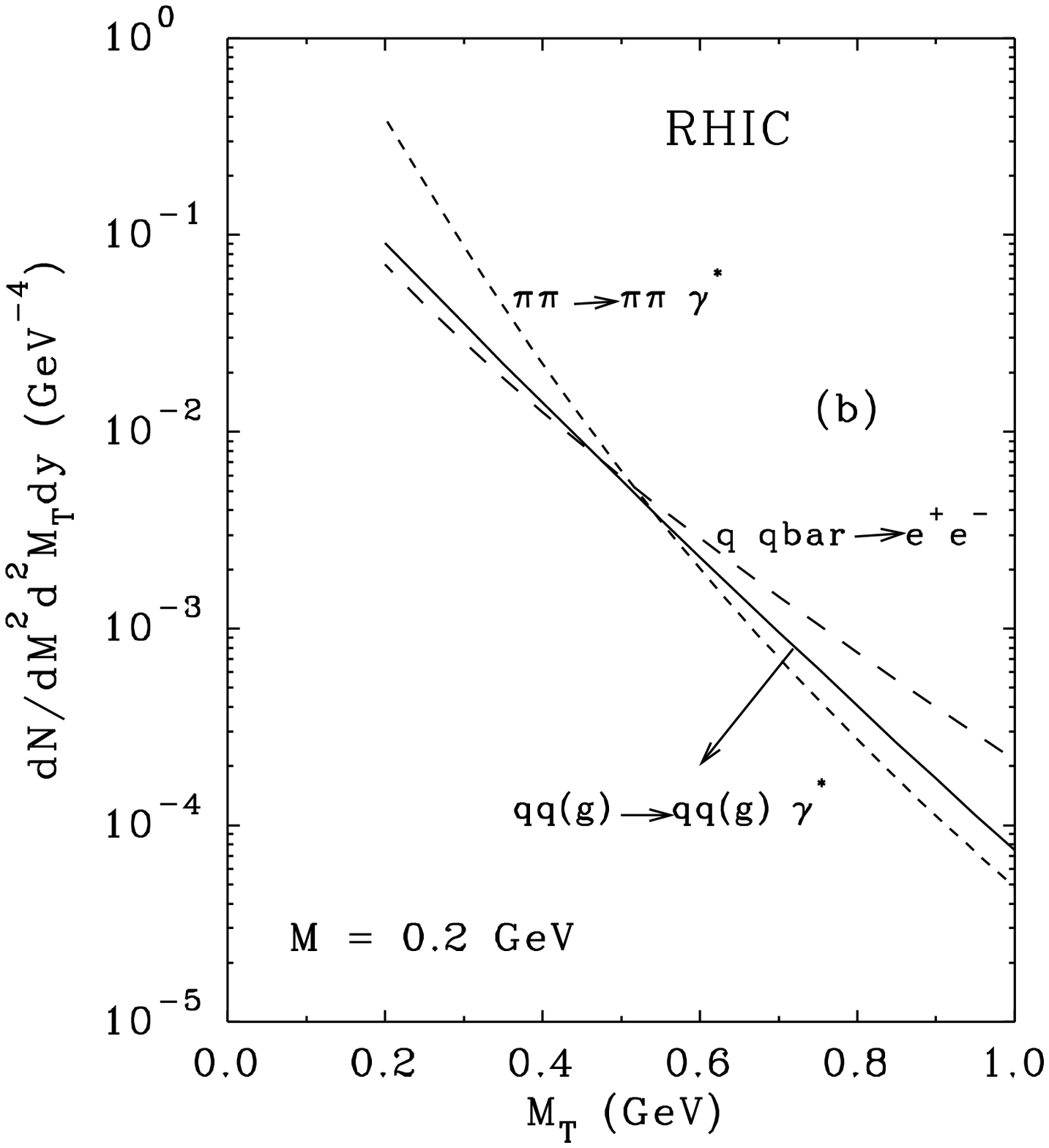,height=2.25in,width=3.25in}
\vskip 0.2cm
\psfig{figure=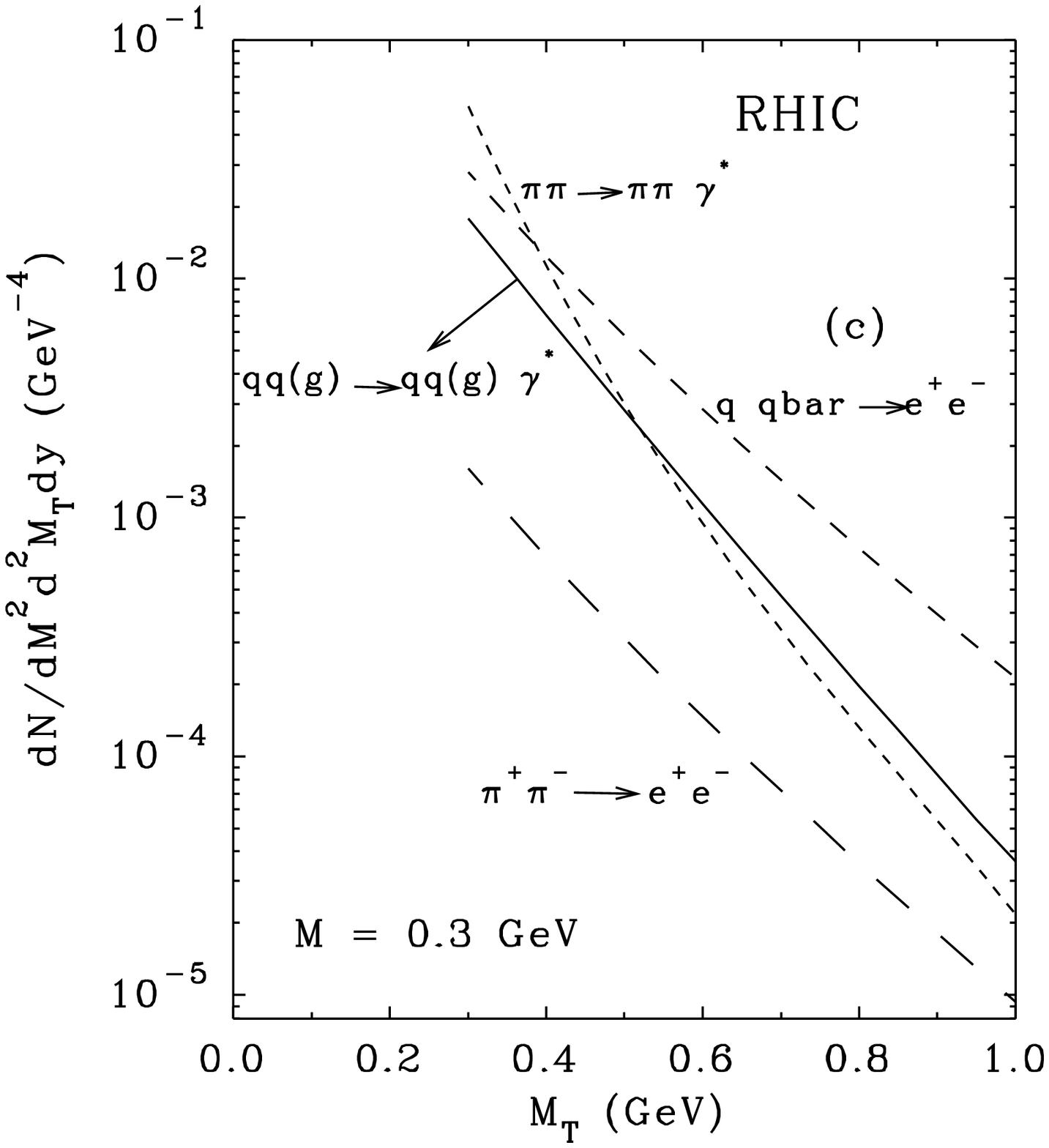,height=2.25in,width=3.25in}
\vskip 0.2cm
\psfig{figure=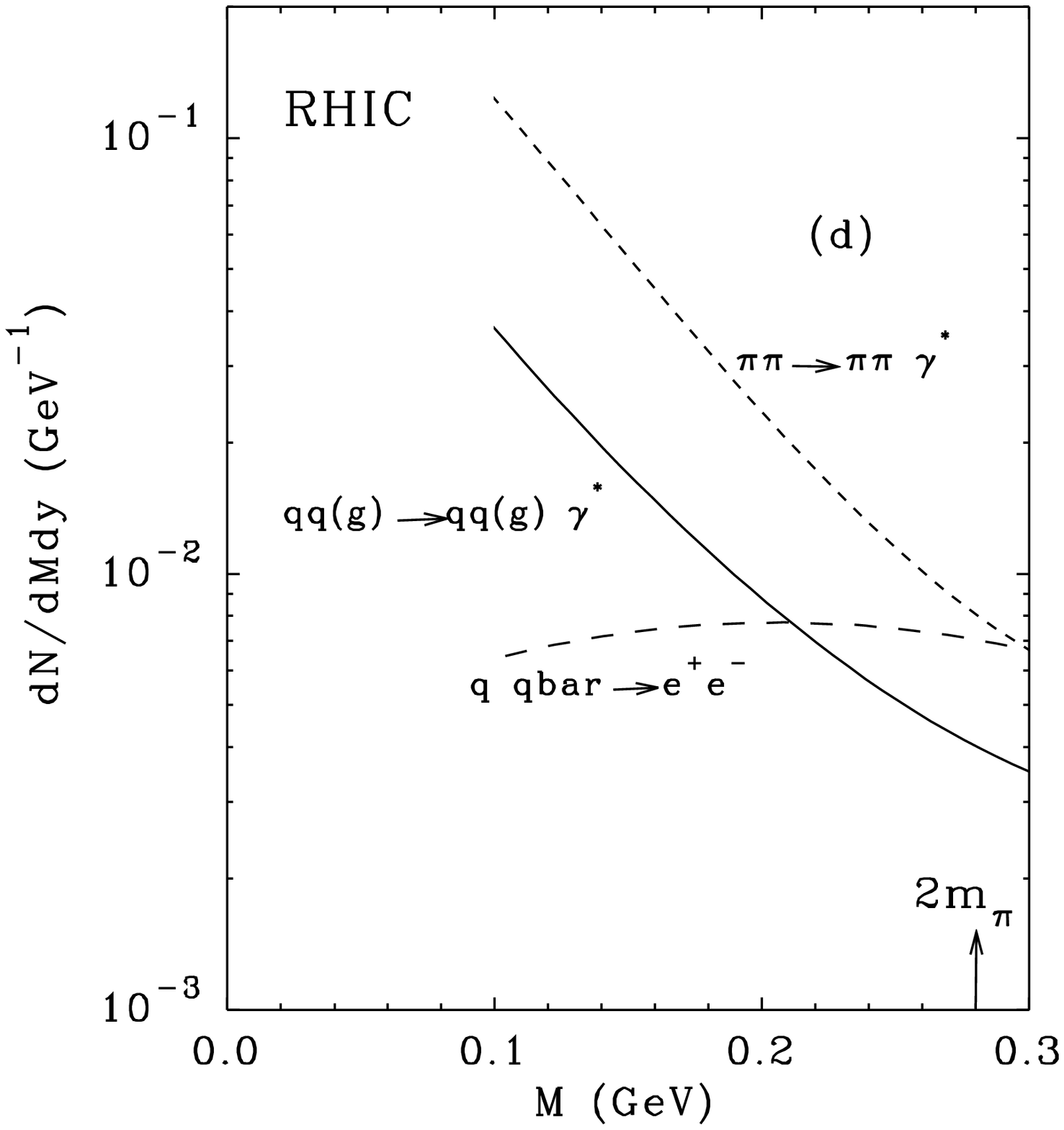,height=2.25in,width=3.25in}
\vskip 0.4cm
\caption{(a--d): Same as fig.~2, for RHIC energies.}
\end{figure}

We have further found (not shown here for reasons of space) that, at 
LHC energies the quark
driven bremstrahlung processes start dominating over the pion driven 
bremstrahlung processes even at relatively smaller $M_T$, as the slopes of
the quark-driven processes are much smaller.
 This aspect remains true even
in the invariant mass spectrum, and the contributions become similar at $M=$
0.3 GeV. 

These results also clearly reveal the rapidly changing
importance of the different processes considered here leading to low mass
dileptons, as the available energy (initial conditions) changes and
as the invariant mass $M$ assumes varying values. When detailed results
are available these considerations may help resolve different contributions.

We envisage an increase by a factor of 2--4 in the dilepton yield as we go from
SPS to RHIC energies, and by a factor of 15-20 as we go from SPS to 
LHC energies.  Thus the existence of a longlived interacting system would be
characterized by an intense glow of low mass dileptons.
This means a large increase in the electromagnetic signals, as compared to
the estimates  done by using only pion and quark annihilations.
 
It is well known that ratios of particle spectra can sensitively reveal the
details of the variations of the underlying processes. We have seen in
Eq.(8) that
the transverse mass-spectra for low mass dileptons are proportional to
$1/M^2$. In figs.4--6  we have 
plotted the  ratio of $M^{2} \,dN/d^{2}M_{T}dM^{2}dy$ at $M =$ 0.1 GeV to 
that for $M =$0.2 GeV and $M =$0.3 GeV both with (solid line) and without 
(dashed line) the transverse flow at SPS, RHIC, and LHC respectively.
We have verified that the (oscillatory) structure seen in the results without 
the transverse flow has its origin in the structure in $\pi \pi$ scattering
cross-section,
which is sampled in the process. One can also show that if there is no
transverse expansion of the system then the ratios as depicted here would
be independent of the initial temperature, that is they would be identical
for SPS, RHIC, and LHC energies, which is also seen from these figures. 
A deviation from this universal behaviour  is indicative of the increasing
importance of the transverse flow as one increases the initial temperature
of the system, which in turn decides the overall life-time of the system.


\begin{figure}
\psfig{figure=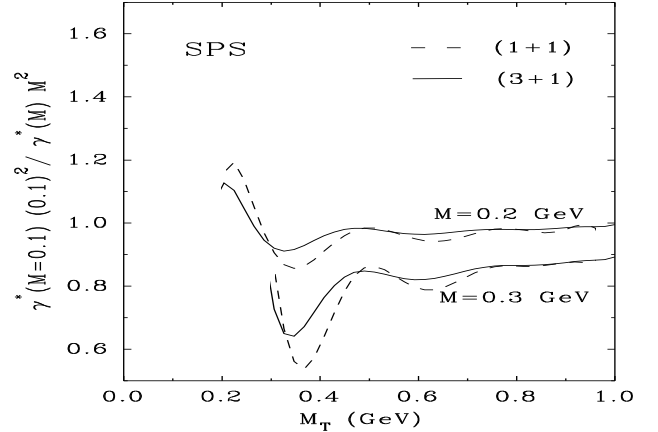,height=2.25in,width=3.25in}
\vskip 0.4cm
\caption{ The ratio of $M^2$ weighted
 differential dielectron yield $M^2 dN/ dM^2d^2M_{T}dy$ at $M =$ 0.1 GeV 
to that at $M =$ 0.2 GeV and $M =$ 0.3
GeV as a function of transverse mass $M_T$ for SPS energies.
 The solid curve gives the total
contribution (quark matter + hadronic matter) with the transverse 
flow. Similarly the dashed curve gives the total contribution without
the transverse flow.}
\end{figure}
\begin{figure}
\psfig{figure=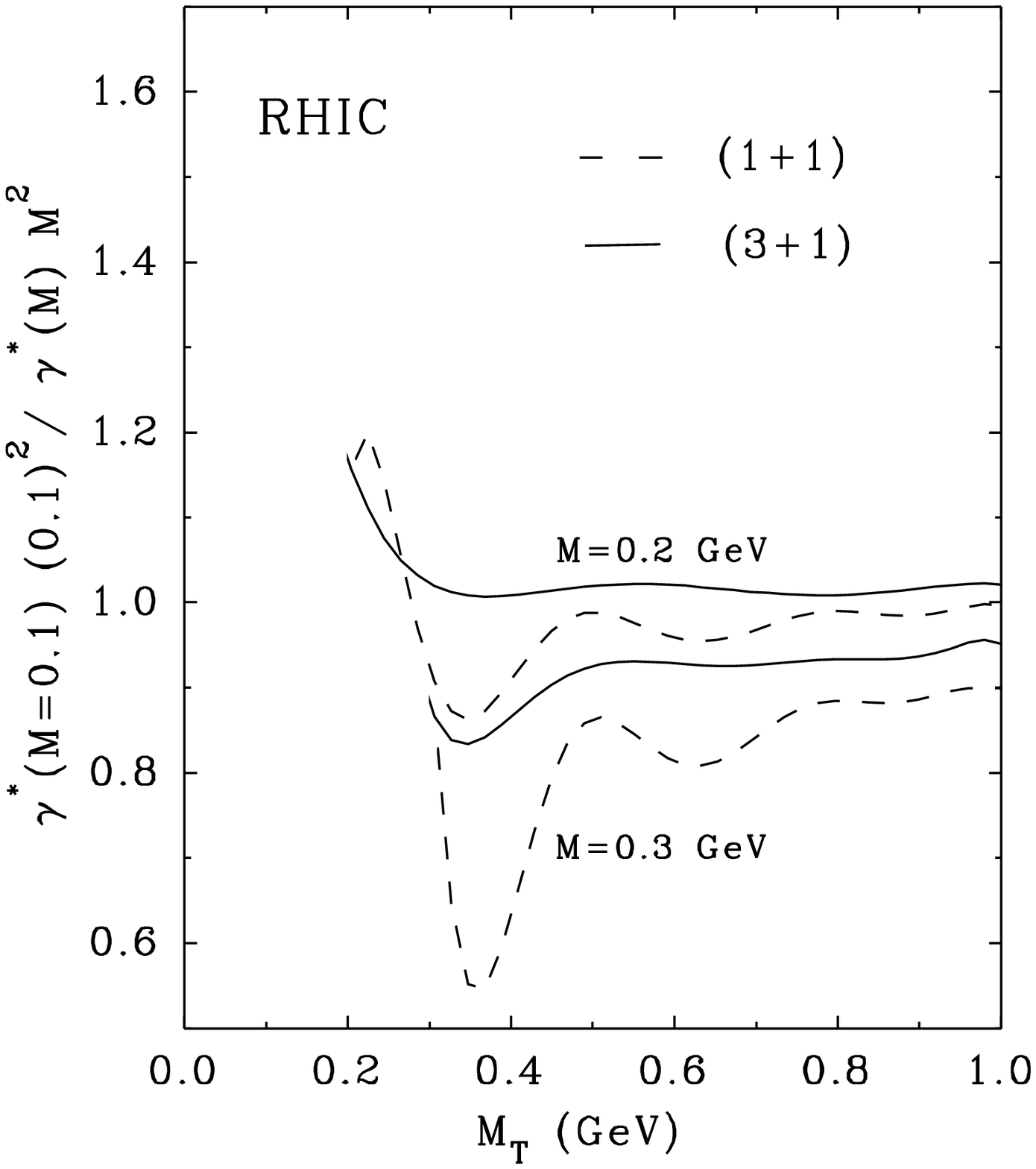,height=2.25in,width=3.25in}
\vskip 0.4cm
\caption{Same as fig.~4 for RHIC energies. The definition of the
solid and the dashed curves are same as in fig.~4.}  
\end{figure}
\begin{figure}
\psfig{figure=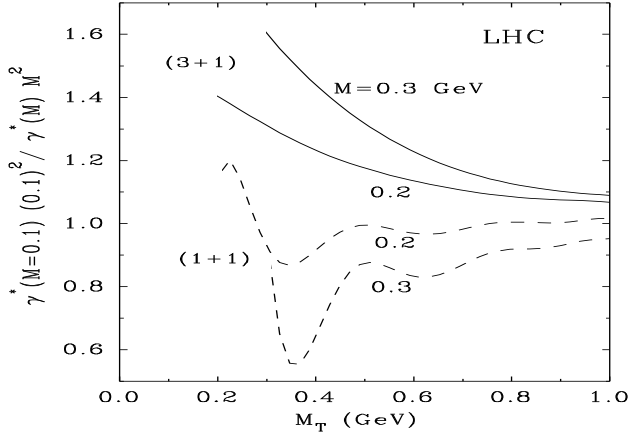,height=2.25in,width=3.25in}
\vskip 0.4cm
\caption{Same as fig.~4 for LHC energies. The definition of the
solid and the dashed curves are same as in fig.~4.}  
\end{figure}


Finally, while investigating the dependence of our results on the
freeze-out temperature  
a successively increasing dependence on this last stage of the
interacting system was seen as we go from SPS to RHIC to LHC energies. 
The largest
sensitivity is thus seen for the LHC energies (see fig.~7)
, where the life-time of the
interacting system is longest, giving the transverse flow effects ample
scope to come into full play.

\begin{figure}
\psfig{figure=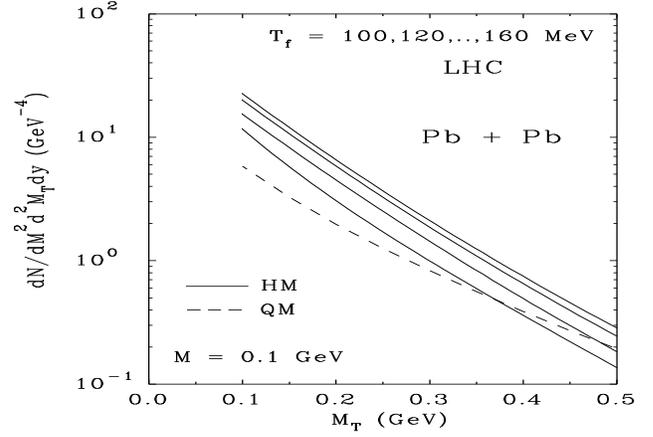,height=2.25in,width=3.25in}
\vskip 0.4cm
\caption{Sensitivity of the low mass dielectron spectra to the
freeze- out temperature at LHC energies for $M =$ 0.1 GeV.}
\end{figure}

\subsection{ Soft Photons}

In a manner similar to the above, we have plotted the rates for 
different photon producing processes at $T=160$ MeV (fig.8).
We see that the quark and pion driven bremsstrahlung processes dominate
upto energies of a few hundred MeV, after which they fall rapidly.

\begin{figure}
\psfig{figure=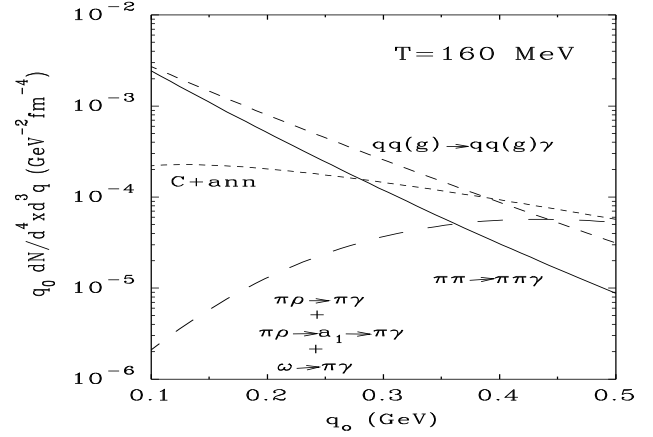,height=2.25in,width=3.25in}
\vskip 0.4cm
\caption{Soft photon production rate at $T =$ 160 MeV  
from quark and pion bremsstrahlung, Compton + annihilation
processes and the sum of the main hadronic reactions as shown in the figure.}
\end{figure}

Space-time integrated results for RHIC energies are
shown in fig.9. We find that soft photons having transverse
momenta of upto a few hundred MeV mostly originate from pion driven
bremsstrahlung processes, and once again the existence of a longlived
interacting system is revealed by an intense glow of soft photons,
once the background of decay photons is removed.

\begin{figure}
\psfig{figure=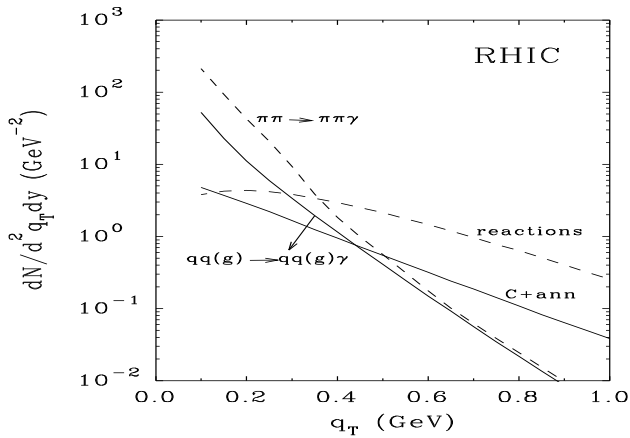,height=2.25in,width=3.25in}
\vskip 0.4cm
\caption{The transverse momentum distribution of soft photons from 
different mechanisms at RHIC energies. The sum of the contribution 
$\pi \rho \rightarrow \pi \gamma$, $\pi \rho \rightarrow a_1 \rightarrow
\pi \gamma$ and the decay $\omega \rightarrow \pi \gamma$ is referred as
{\it 'reactions'}.}
\end{figure}
%
Even though the relative importance of the various contributions was
found to be similar at SPS and LHC energies, we envisage an increase 
by a factor of 2--4 in the yield
of photons having $p_T=$ 200 MeV, as we go from SPS to RHIC and an increase
by a factor of almost 10 as we go from SPS to LHC energies.  The
complete dominance of soft photons in determining the multiplicity
of photons produced is seen from fig.10. Note that we have included only
photons having $p_T>$ 100 MeV, for this discussion, as we know that the
yield for lower $p_T$ is subject to Landau Pomeranchuk effect.
It may be noted, however, that unlike the case of dileptons, the contribution
of  the reaction $\pi \pi \rightarrow \rho \gamma$ which is equivalent
to $\pi \pi \rightarrow \pi \pi \gamma$ was included in the esimates of
single photons \cite{Kapusta}, and thus the increase above does not 
necessarily mean a new source. It merely points to a rapid rise in the
yield of single photons having $p_T < $ 300--400 MeV.
\begin{figure}
\psfig{figure=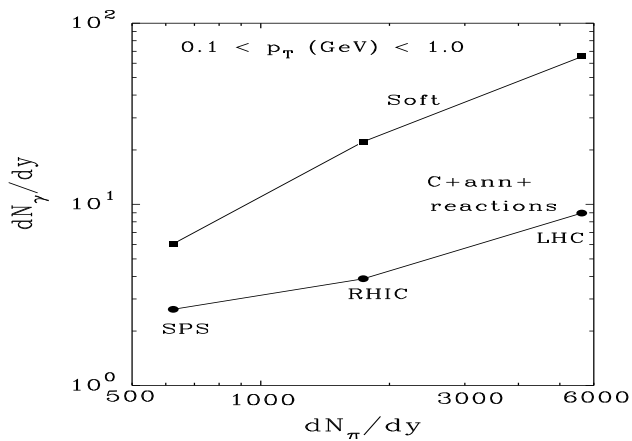,height=2.25in,width=3.25in}
\vskip 0.4cm
\caption{Soft photons vs. photons from Compton plus annihilation
processes from the QGP and hadronic reactions at SPS, RHIC, and LHC energies
from central collision of two lead nuclei.} 
\end{figure}

\section{DISCUSSION}

There are a number of aspects which should be discussed before we draw our
conclusions. We have already discussed the validity of the
soft-photon approximation in sect. II.

It is well-known that bremsstrahlung radiation could be very large for
light particles and one may worry about this aspect for the radiations
from the quark matter. We have, however, used the thermal mass of quarks
while evaluating the $d\sigma/dt$ as well as the kinematics of the collision,
which is appropriate for fermions moving in a hot medium. We have already
stated that if we extend this argument to quark 
annihilation as well, then the bremsstrahlung processes become the leading
contributors to this mass range for dileptons.

Any study of soft electromagnetic radiations must address the question 
of Landau- Pomeranchuk \cite{Landau} suppression of 
such processes in a dense medium.
The Landau- Pomeranchuk effect provides that if the formation time of a
particle is more than the time between two collisions, the emission of
the particle could be considerably suppressed due to destructive
interference of multiple scatterings. We would like to draw the
attention of the readers to arguments developed in Ref. \cite{pradip}
earlier, about the extent of the modifications to our predictions due to this.
It can be argued that our results
for the sum of the radiations from the interacting system will remain 
fairly free from the effects of Landau- Pomeranchuk suppression
till we restrict ourselves to photons and dileptons having energies 
larger than a few hundred MeV. The Landau- Pomeranchuk suppression
could be severe  for lower energies as, indeed, 
demonstrated by Cleymans et al. \cite{jean}.

These considerations have an interesting connotation. Recall that
we have taken the initial time as 1 fm/c. It is quite likely that the
QGP may be thermalized much more quickly \cite{kms}, and then we can have
a larger initial temperature for the same multiplicity of the
particles. We have seen earlier that the partonic density can then be
much higher, and thus there would be a suppression of soft radiations.
Thus our choice of $\tau_{i} = $1 fm/c ensures that we start our 
evaluations {\it after} the Landau- Pomeranchuk effect has lost its 
dominating effect, and that our estimates remain reasonable.
 A more complete treatment will include the Landau
Pomeranchuk effect and thus these suppressions would be automatically,
and more properly accounted for.

We have approximated the hadronic phase as a non-interacting gas of $\pi$,
$\rho$, $\omega$, and $\eta$ mesons. How will the results differ for
a richer hadronic matter, which would result in a reduced life-time for the
mixed-phase? A richer equation of state for hadornic matter will also imply
a smaller speed of sound, and the attendant slower cooling of the system,
and  a longer life time for the hadronic phase. Thus, it was found recently
that the results for single photons \cite{cape} with a
truncated equation of state
as used here and a resonance gas containing all hadrons, for the 
hadronic matter left the final results essentially unaltered. Similar results
should be expected here, due to the similarity of the rates for the
quark and pion driven processes (see. fig.1).

All our evaluations are made with the assumption that the QGP, as produced
initially, is in kinetic and chemical equilibrium, and  that its evolution is
isentropic. It is quite likely that the plasma as produced in relativistic
heavy ion collisions is neither in kinetic nor in chemical equilibrium.
How will this affect our findings? Even though the kinetic equilibrium
could be achieved quickly enough, the chemical equilibration itself may 
not be achieved at all \cite{klaus,biro,mustafa}. The contributions of the
QGP part is then easily obtained by introducing the products $\lambda_i
\lambda_j$, where $\lambda_i$ is the fugacity of the parton species $i$,
in our expressions \cite{strik}. 
Needless to add that the overall contribution could
come down by a factor of upto 10 or more depending upon the initial
conditions. So far there is no treatment which could model the
hadronization of QGP which is far from  chemical equilibrium. It is not
even clear that such a matter will go through a mixed phase. The description
of the hadronic phase (if any) also gets uncertain. However, a very
interesting outcome of this scenario could be a complete absence of radiations
from  the hadronic processes, if the QGP phase is not followed by an interacting
hadronic matter living for some finite time! This could be of great interest.

What could be other sources of low mass dileptons? It was suggested some 
time ago \cite{ssg} that $\pi \rho \rightarrow \pi\, e^+e^-$ could contribute
to low mass dileptons. This has now been evaluated \cite{kevin2}, 
and it is found
to contribute less than the bremsstrahlung processes at lower masses. However
the bremsstrahlung contribution decreases rapidly and for  $M>$ 300 MeV,
and the above reaction contributes at a level of 10--50\% of the pionic annihilation.
It will be of interest to study the transverse mass distribution of this 
reaction, as it is likely to be different. 

\section{SUMMARY}

We have calculated the transverse mass distribution of low mass
dileptons and transverse momentum distribution of soft photons
from central collision of two lead nuclei at CERN SPS, BNL RHIC,
and CERN LHC energies. We assume that the collision leads to a
thermalized and chemically equilibrated quark gluon plasma at the
proper time $\tau_{i} = $1 fm/c. The plasma then expands, cools,
and gets into a mixed phase at $T = $160 MeV. After all the quark
matter is adiabatically converted to hadronic matter, it cools again, 
and undergoes a freeze-out at $T = $140 MeV. We have considered a 
boost invariant longitudinal and cylindrically symmetric transverse
expansion. This is, to our knowledge, the first treatment of the 
dynamics of soft electromagnetic radiations in such collisions, 
with transverse expansion, whose effect is seen to be large when
the life-time of the interacting system is large.

We find that the formation of such a system may be characterized 
by an intense glow of soft electromagnetic radiations, whose
features depend sensitively on the last stage of evolution, once we remove the
background of decay photons or dileptons.

We are grateful to  Hans Eggers and Kevin Haglin for very many useful
 discussions during the course of this work.

\section*{Figure Captions}

Figure~(1a--c): The production rate of low mass dielectrons from quark
and pion bremsstrahlung at $T = $ 160 MeV.
In addition, the contribution of quark annihilation process is given
for a comparison. These results are shown for (a) $M =$ 0.1 GeV, (b)
 $M =$ 0.2 GeV, and (c) $M =$ 0.3 GeV respectively.

\vskip 0.25in

Figure~(2a--d): The transverse mass distribution of low mass dielectrons
at SPS energies including bremsstrahlung process and annihilation 
process in the quark matter and the hadronic matter. We give the results
for invariant mass M equal to
 0.1 GeV (a), 0.2 GeV (b), and  0.3 GeV (c) 
respectively. The invariant mass distribution of low mass
dielectrons are also shown (d).

\vskip 0.25in

Figure~(3a--d): Same as fig.~2, for RHIC energies.

\vskip 0.25in

Figure~4: The ratio of $M^2$ weighted
 differential dielectron yield $M^2 dN/ dM^2d^2M_{T}dy$ at $M =$ 0.1 GeV 
to that at $M =$ 0.2 GeV and $M =$ 0.3
GeV as a function of transverse mass $M_T$. The solid curve gives the total
contribution (quark matter + hadronic matter) with the transverse 
flow. Similarly the dashed curve gives the total contribution without
the transverse flow.

\vskip 0.25in

Figure~5: Same as fig.~4 for RHIC energies. The definition of the
solid and the dashed curves are same as in fig.~4.  

\vskip 0.25in

Figure~6: Same as fig.~4 for LHC energies. The definition of the
solid and the dashed curves are same as in fig.~4.  

\vskip 0.25in

Figure~7: Sensitivity of the low mass dielectron spectra to the
freeze- out temperature at LHC energies for $M =$ 0.1 GeV.

\vskip 0.25in

Figure~8: Soft photon production rate at $T =$ 160 MeV  
from quark and pion bremsstrahlung, Compton + annihilation
processes and the sum of the main hadronic reactions as shown in the figure.

\vskip 0.25in

Figure~9: The transverse momentum distribution of soft photons from 
different mechanisms at RHIC energies. The sum of the contribution 
$\pi \rho \rightarrow \pi \gamma$, $\pi \rho \rightarrow a_1 \rightarrow
\pi \gamma$ and the decay $\omega \rightarrow \pi \gamma$ is referred as
{\it 'reactions'}.

\vskip 0.25in

Figure~10: Soft photons vs. photons from Compton plus annihilation
processes from the QGP and hadronic reactions at SPS, RHIC, and LHC energies
from central collision of two lead nuclei. 
\end{document}